\documentclass[journal,fleqn]{IEEEtran}
\IEEEoverridecommandlockouts

\usepackage{amsmath,amssymb,amsfonts,mathtools,bm}
\IfFileExists{bbm.sty}{\usepackage{bbm}\newcommand{\indic}{\mathbbm{1}}}{\newcommand{\indic}{\mathbb{1}}}
\usepackage{cases}

\usepackage{array}
\usepackage{booktabs}
\usepackage{multirow}
\usepackage{makecell}
\usepackage{tabularx}
\newcommand{\cmark}{\ding{51}}
\newcommand{\xmark}{\ding{55}}
\usepackage{algorithm}
\usepackage[noend]{algpseudocode}

\usepackage{graphicx}
\usepackage{subcaption}
\usepackage{comment}
\usepackage{xcolor}
\usepackage{url}
\usepackage{textcomp}
\usepackage{cite}
\usepackage{pifont}      
\usepackage{caption}

\usepackage[colorlinks=true,
            citecolor=blue,
            urlcolor=blue,
            linkcolor=blue,
            bookmarks=false]{hyperref}

\newcommand{\E}{\mathbb{E}}
\DeclareMathOperator{\PPOLocalOp}{\mathcal{U}_{\mathrm{PPO}}}  
\newcommand{\Ld}{D}   

\begin{document}

\title{Byzantine-Resilient Federated Multi-Agent Optimization Framework for Cyber-Secure Interconnected Microgrids}

\author{
Ali~Peivand\,,
and Seyyed~Mostafa~Nosratabadi\,,~\IEEEmembership{Member, IEEE}

}


\maketitle

\begin{abstract}
The escalating digitalization of distribution networks has exposed interconnected Microgrid (MG) clusters to Stealthy False Data Injection Attacks that bypass Bad Data Detectors and propagate through tie-line couplings and shared learning channels. This paper proposes BR-FedMAPPO, a Byzantine-Resilient Federated Multi-Agent Proximal Policy Optimization framework that learns a triple-surface Moving Target Defense and an adaptive isolation strategy for cyber-secure operation. Each MG hosts a local Actor-Critic Agent whose policy is partitioned into a globally federated shared encoder and a privately retained action head, so no MG exposes the configurations, cardinality, or locations of its D-FACTS lines, Battery Energy Storage (BES) units, or tie-line capacities. The action vector perturbs D-FACTS reactances, redirects BES injections, reshapes inter-MG exchanges, and includes a continuous islanding signal. A two-stage Byzantine-resilient aggregation rule combines trimmed-mean filtering with reward-weighted updates. This scheme incorporates a detection-quality score based on the F1-score and False Positive Rate to penalize clients causing false alarms. Simulation results on four interconnected MGs based on the IEEE 30- and 118-bus test systems demonstrate effective mitigation of coordinated S-FDI attacks, containment of cascading disruptions through adaptive isolation, and protection of distributed learning channels against malicious model manipulations while maintaining cost-aware dispatch performance.
\end{abstract}


\section{Introduction}
\IEEEPARstart{T}{he} rapid digitalization of modern power systems has transformed traditional distribution networks into complex cyber-physical infrastructures dependent on open communication channels \cite{rref1,ref2,ref3}. While this evolution facilitates the integration of renewable energy and demand response, it simultaneously introduces severe vulnerabilities to cyber threats \cite{ref4,ref5}. Among these, Stealthy False Data Injection Attacks (S-FDIAs) pose a critical risk: attackers craft malicious injections that manipulate state estimation without triggering conventional Bad Data Detectors (BDDs) \cite{ref6}. By subtly altering operational data, adversaries can drive power dispatch, protection, and market operations into unsafe states, with severe physical and economic consequences that motivate defensive strategies beyond traditional perimeter security.

The sophistication of S-FDIAs has rendered classical detection schemes inadequate. Early attack models assumed static grid configurations, but recent work shows that adversaries can bypass even advanced deep-learning detectors by training surrogate models on historical grid data \cite{ref7,ref8}. The threat multiplies within interconnected MG clusters, where a single compromised MG can propagate corrupted measurements through shared tie-lines and collaborative control algorithms \cite{ref9}. Because conventional anomaly detectors target random sensor noise rather than strategically crafted injections, these systems require proactive defenses that dynamically alter the attack surface instead of relying solely on reactive detection \cite{ref27,ref28}.

Moving Target Defense (MTD) has emerged as a promising proactive countermeasure by intentionally modifying physical or cyber parameters of the grid \cite{ref10,ref11}. The canonical implementation uses distributed flexible AC transmission system devices to perturb line reactances \cite{ref12}, dynamically shifting the measurement Jacobian and forcing pre-calculated attack vectors out of the valid state space so they are exposed to standard residual tests. Subsequent studies extended this concept to hidden defense strategies \cite{ref13}, optimal device placement \cite{ref14}, noise-robust configurations \cite{ref16}, event-triggered switching \cite{ref17}, and stealth-aware cost optimization \cite{ref18}. Despite these advances, most existing strategies rely on centralized control that violates the privacy constraints of multi-owner MG clusters, and they conventionally perturb only network reactance, neglecting other control surfaces such as storage dispatch or dynamic inter-grid exchanges \cite{ref15}.

To address adaptivity and privacy, recent literature has explored the intersection of Deep Reinforcement Learning (DRL) and Federated Learning (FL) for grid defense \cite{ref19,ref20,ref21}. Multi-Agent Proximal Policy Optimization (MAPPO) lets distributed agents learn cooperative defensive actions with decentralized execution \cite{ref22,ref23,ref24}, while FL lets independent operators train a global policy without sharing raw data \cite{ref25,ref26}. However, critical vulnerabilities persist: standard aggregation rules such as federated averaging are notoriously susceptible to Byzantine faults, where even a single compromised participant can derail learning \cite{ref29,ref30}. Moreover, current federated DRL approaches typically assume homogeneous action spaces and ignore the physical coupling between MGs, leaving adaptive islanding and topology privacy unaddressed \cite{ref31,ref32,ref33}.

Despite individual advances, a comprehensive framework addressing the intricate cyber-physical interactions in networked MGs is lacking. Centralized defenses are incompatible with the privacy requirements of autonomous MGs, and DRL approaches have not integrated multi-surface physical defenses with Byzantine-resilient federated aggregation. Structural heterogeneity, Differential Privacy (DP) of local control topologies, and autonomous adaptive isolation remain largely unexplored in tandem. Table \ref{tab:qualitative_compact} provides a qualitative comparison with recent cyber-secure MG control studies, illustrating the absence of a unified framework capable of securing decentralized grids against coordinated cyber-physical manipulation.

\begin{table}[!t]
\caption{Compact qualitative comparison with recent cyber-secure MG control frameworks.}
\label{tab:qualitative_compact}
\centering
\tiny
\renewcommand{\arraystretch}{0.94}
\setlength{\tabcolsep}{1.05pt}

\resizebox{\columnwidth}{!}{%
\begin{tabular}{@{}l@{\hspace{2pt}}c@{\hspace{2pt}}c@{\hspace{2pt}}c@{\hspace{2pt}}c@{\hspace{2pt}}c@{\hspace{2pt}}c@{\hspace{2pt}}c@{\hspace{2pt}}c@{}}
\toprule
\textbf{Ref.} &
\textbf{MG} &
\textbf{MTD} &
\textbf{Alg.} &
\textbf{FL} &
\textbf{BR} &
\textbf{Atk.} &
\textbf{Priv.} &
\textbf{Det.} \\
\midrule

\cite{ref8}  & S    & R-only & H-MTD+NN      & \xmark & \xmark & \xmark & --       & NN-pool \\
\cite{ref16} & S    & R-only & R-MTD(QP)     & \xmark & \xmark & \xmark & --       & $\chi^2$/KL \\
\cite{ref17} & S    & R-only & ET-MTD        & \xmark & \xmark & \xmark & --       & D+P \\
\cite{ref18} & S    & R-only & S-MTD         & \xmark & \xmark & \xmark & --       & UML \\
\cite{ref21} & MGC  & None   & Fed-DL        & \cmark & \xmark & \xmark & --       & Disc./FDI \\
\cite{ref24} & MGC  & None   & MAPPO         & \xmark & \xmark & \xmark & --       & None \\
\cite{ref25} & VPPC & None   & Fed-DRL       & \cmark & \xmark & \xmark & --       & None \\
\cite{ref26} & MGC  & None   & F-MARL        & \cmark & \xmark & \xmark & DP       & None \\
\cite{ref28} & SBS  & None   & SecFedDL      & \cmark & \xmark & \xmark & HE+DP    & FDI-cls \\
\cite{ref31} & SBS  & None   & VFL           & \cmark & \xmark & \xmark & DP       & FDI-cls \\
\cite{ref32} & Gen. & None   & D-FL          & \cmark & TM-like & \xmark & --       & None \\
\cite{ref33} & NMG  & None   & VFRL          & \cmark & \xmark & scripted & None     & None \\

\midrule
\textbf{This paper} &
\textbf{NMGC} &
\textbf{Triple} &
\textbf{Fed-MAPPO+HH} &
\cmark &
\textbf{\cmark~TM+RW} &
\textbf{\cmark~RL-act.} &
\textbf{DP+SP} &
\textbf{4-stat 2/4} \\
\bottomrule
\end{tabular}%
}

\vspace{-1.1mm}
\begin{minipage}{\columnwidth}
\scriptsize
\end{minipage}

\end{table}

In Table \ref{tab:qualitative_compact}, the abbreviations S, MGC, VPPC, SBS, Gen, NMG, and NMGC denote single systems, MG clusters, virtual power plant clusters, single bus-set benchmarks, generic systems, networked MGs, and networked MG clusters, respectively. The security columns encompass Byzantine Resilience (BR), Trimmed-Mean filtering (TM), Reward-Weighted updates (RW), homomorphic encryption (HE), DP, and Structural Privacy (SP). This paper addresses the identified gaps through the following contributions:
\begin{itemize}
\item \textbf{Byzantine-Resilient Framework}: We propose a Byzantine-resilient federated multi-agent proximal policy optimization framework that enables a triple-surface MTD utilizing reactance control, BES dispatch, and tie-line exchange, coupled with adaptive islanding as a learned control action.
\item \textbf{Federated Architecture}: A shared-encoder and private-action-head federated architecture allows representation sharing across MGs while maintaining local topology-specific action heads, preserving the structural privacy of defense device placements.
\item \textbf{Novel Reward-Weighted Aggregation Rule}: We introduce a reward-weighted Byzantine aggregation rule that evaluates participant influence based on episodic reward and detection quality, providing superior robustness against adversarial model poisoning compared to standard averaging or median-based filters.
\item \textbf{Autonomous Adaptive Isolation}: Operating concurrently with the cost-aware multi-surface MTD, the framework integrates an intelligent islanding mechanism governed by a multi-statistic voting detector. Upon identifying a critical compromise, the local agent isolates the affected MG and employs continuous hysteresis control to maintain the disconnection, preventing physical and cyber propagation across tie-lines until the threat is neutralized.
\end{itemize}

The remainder of this paper is organized as follows. Section \ref{sec:framework} describes the proposed framework. Section \ref{sec:setup} presents the experimental setup. Section \ref{sec:results} provides simulation results and ablation studies. Section \ref{sec:conclusion} concludes the paper.
\section{Proposed BR-FedMAPPO Framework}\label{sec:framework}

\begin{figure*}[!t]
\centering
\includegraphics[width=0.92\textwidth]{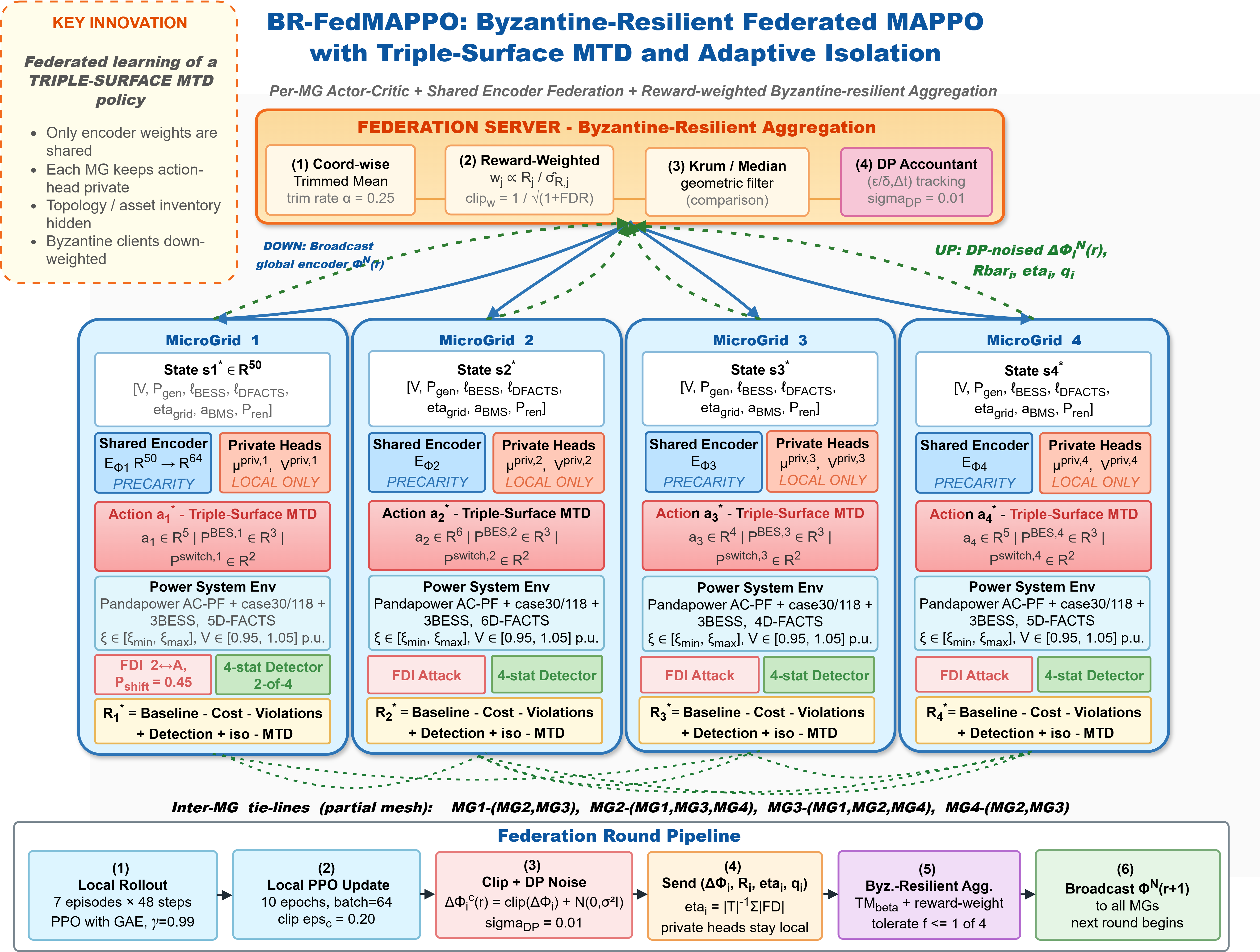}
\caption{Overall architecture of the proposed BR-FedMAPPO framework.}
\label{fig:architecture}
\end{figure*}

This section presents the proposed BR-FedMAPPO framework. A schematic overview is provided in
Fig.~\ref{fig:architecture}, and the complete pseudo-code is given in
Algorithm~\ref{alg:brfedmappo}.

As shown in Fig.~\ref{fig:architecture}, each MG hosts a local MAPPO
agent with a shared encoder and a private actor/critic head. The
triple-surface MTD action space jointly perturbs line reactances, BES
dispatch, and inter-MG power exchanges. The federation server applies
a two-stage Byzantine-resilient aggregation rule combining
coordinate-wise trimmed-mean filtering with a reward-weighted update;
differential privacy is enforced on encoder updates before
transmission.

\subsection{Mathematical Modelling of the Interconnected MGs}

The framework features a partially meshed cluster of $M$ MGs
whose internal dynamics are governed by a nonlinear AC power-flow
model that determines voltage magnitudes $V_{i,n}^{t}$ and angles
$\theta_{i,n}^{t}$ at each control step. The active- and
reactive-power injections at each bus are computed from the standard
AC power-flow equations as in \cite{ref12}. Each MG owns a bank
$\mathcal{L}_i^{D}\subseteq\mathcal{L}_i$ of D-FACTS-equipped lines
whose reactances $x_{l}^{t}=x_{l}^{0}+\Delta x_{l}^{t}$ are perturbed
within $[\Delta x_{\min},\Delta x_{\max}]$ by the agent action, and a
set $\mathcal{B}_i$ of BES units with charge/discharge powers
$P_{ch,i,k}^{t},P_{dis,i,k}^{t}$ whose residual energy evolves as:
\begin{equation}\label{eq:bes}
E_{i,k}^{t+1}=E_{i,k}^{t}+\left(\eta_{c}\,P_{ch,i,k}^{t}
-\frac{P_{dis,i,k}^{t}}{\eta_{d}}\right)\Delta t
\end{equation}
subject to $E_{i,k}^{t}\in[E_{\min},E_{\max}]$ and
$|P_{\cdot,i,k}^{t}|\le P_{\max}^{BES}$. Conventional generation,
voltage, and line-loading limits are imposed at every bus and line,
$P_{gen,i,g}^{\min}\!\le\!P_{gen,i,g}^{t}\!\le\!P_{gen,i,g}^{\max}$,
$V_{\min}\!\le\!V_{i,n}^{t}\!\le\!V_{\max}$, and
$S_{i,l}^{t}\!\le\!S_{i,l}^{\max}$. When MG~$i$ is islanded, all
tie-line flows are forced to zero, $P_{ij}^{exch,t}=0$ for
$\forall j\in\mathcal{N}_i$, and the power balance is augmented with
a load-shedding term:
\begin{equation}\label{eq:balance}
\sum_{g}P_{gen,i,g}^{t}+\sum_{k}P_{BES,i,k}^{t}
=\sum_{l}\!\bigl(P_{load,i,l}^{t}-P_{shed,i,l}^{t}\bigr)+P_{loss,i}^{t}
\end{equation}
with $0\le P_{shed,i,l}^{t}\le P_{load,i,l}^{t}$. The shed power
carries a high penalty in the reward, so islanding is invoked only
when isolating an ongoing FDI propagation outweighs the cost of
unserved energy.

\subsection{False Data Injection Attack Model}

The attacker is an external adversary who has compromised a subset of
measurement channels on MG~$i$ and injects an additive vector
$c_{i}^{t}$ into the observation:
\begin{equation}\label{eq:fdi}
\tilde{z}_{i}^{t}=z_{i}^{t}+c_{i}^{t}
\end{equation}
where $z_{i}^{t}$ is the noisy real measurement. Two attack types
are considered. In the sparse type, $c_{i}^{t}$ has only a
fraction $\rho_{atk}$ of non-zero entries with magnitudes
proportional to the local measurement. In the stealthy type,
$c_{i}^{t}=H_{i}\,a_{S\text{-}FDI,i}^{t}$ lies in the column space of
the linearized Jacobian $H_i$, which is provably invisible to a
chi-squared bad-data test that assumes the same $H_i$. Attacks arrive
as temporally coherent bursts: a fresh burst is triggered with
probability $p_{atk}$ and persists for a random duration in
$[L_{\min},L_{\max}]$ steps, with the attack vector regenerated each
step to track the operating point.
Each MG hosts an unattended detector that combines four statistically
distinct tests with a configurable voting rule: (i) a
normalized-magnitude residual
$\|\tilde{z}_{i}^{t}-\mu_{i}\|/\|\mu_{i}\|$ relative to a calibration
mean $\mu_{i}$; (ii) a weighted Mahalanobis statistic with an
adaptive per-channel scale
$\sigma_{i,m}\!=\!\max(\hat{\sigma}_{i,m},\beta\mu_{i,m},\sigma_{\min})$
that prevents the variance collapse breaking plain chi-squared tests;
(iii) the reconstruction error
$e_{AE,i}^{t}=\|\tilde{z}_{i}^{t}-f_{AE,i}(\tilde{z}_{i}^{t})\|^{2}$
of a local autoencoder trained on the same calibration buffer; and
(iv) a cumulative-sum change-point statistic that captures slow
drifts. Each test fires when its statistic exceeds a data-driven
threshold calibrated from an empirical percentile of the calibration
buffer. The final flag $d_{i}^{t}\!\in\!\{0,1\}$ is set when the
number of triggered tests reaches a voting threshold $\tau_{vote}$.
The detector also exposes a continuous anomaly score
$\eta_{i}^{t}\!\in\![0,1]$ to the agent as part of the state.

\subsection{Problem Formulation as a Markov Game}

Because each MG observes only its own state, controls only its own
actuators, and receives a partially shared reward, the cooperative
defense problem is a partially-observable Markov game
$\mathcal{G}=\langle\mathcal{F},\mathcal{O}_i,\mathcal{A}_i,
\mathcal{P},r_i,\gamma\rangle$, where $\mathcal{P}$ is the joint
transition kernel induced by the AC-PF solver and FDI process. The
local observation of MG~$i$ is
\begin{equation}\label{eq:obs}
s_{i}^{t}=\bigl[V_{i}^{t},\,p_{g,i}^{t},\,SOC_{i}^{t},\,\Ld_{i}^{t},\,
\eta_{i}^{t},\,I_{i}^{iso,t},\,p_{tie,i}^{t}\bigr]^{\!\top}
\end{equation}
The joint policy maximizes the discounted cumulative team reward
$J^\Theta=\E[\sum_{t}\gamma^{t}\sum_{i}r_{i}^{t}]$ subject to the
constraints above. Training follows the CTDE paradigm. The reward of
MG~$i$ combines a baseline term, operational-cost increment,
voltage/overload penalties, an MTD-activity penalty, a load-shedding
penalty, a BES degradation cost, an islanding penalty, and
detection-related bonuses/penalties:
\begin{equation}\label{eq:reward}
\begin{aligned}
r_{i}^{t}={}&R_{base}-w_{cost}\,\Delta C_{i}^{t}
            -w_{volt}\,n_{viol,i}^{v,t}
            -w_{ol}\,n_{viol,i}^{ol,t}\\
           &-w_{ol,sev}\,\Phi_{i}^{t}-w_{loss}\,P_{loss,i}^{t}
            -w_{mtd}\,\Psi_{i}^{t}-w_{shed}\,P_{shed,i}^{t}\\
           &-w_{iso}\,I_{i}^{iso,t}
            +b_{det}\,d_{i}^{t}\,\indic[\,c_{i}^{t}\neq 0\,]
            -p_{fa}\,d_{i}^{t}\,\indic[\,c_{i}^{t}=0\,]
\end{aligned}
\end{equation}
where $\Psi_{i}^{t}$ aggregates the normalized D-FACTS perturbation,
$\Phi_{i}^{t}$ aggregates the over-100\% line excess, and
$\indic[\cdot]$ is the indicator function. The last two terms reward
correct detection of an active attack and penalize false alarms, so
the agent has a direct incentive to act only when warranted.

\subsection{Action Space Design---Multi-Surface MTD}

A key feature of the proposed action space is its triple-surface
structure. The action vector of agent $i$ is partitioned as:
\begin{equation}\label{eq:action}
a_{i}^{t}=\bigl[\Delta x_{i}^{t},\,p_{BES,i}^{t},\,p_{tie,i}^{t},\,
\alpha_{iso,i}^{t}\bigr]^{\!\top}
\end{equation}
The block
$\Delta x_{i}^{t}\in[\Delta x_{\min},\Delta x_{\max}]^{|\mathcal{L}_i^{D}|}$
is the D-FACTS reactance perturbation and
$p_{BES,i}^{t}\in[-p_{BES}^{\max},p_{BES}^{\max}]^{|\mathcal{B}_i|}$
is the BES dispatch. Also, 
$p_{tie,i}^{t}\in[-p_{tie}^{\max},p_{tie}^{\max}]^{|\mathcal{N}_i|}$
is the inter-MG exchange request and $\alpha_{iso,i}^{t}\in[0,1]$ is
the continuous islanding signal. The binary indicator
$I_{i}^{iso,t}$ follows a hysteresis rule that engages islanding when
$\alpha_{iso,i}^{t}>\tau_{iso}$ and requires a minimum hold of
$T_{iso,\min}$ steps before reconnection. After D-FACTS perturbation,
\begin{equation}\label{eq:Hpert}
H_{i}^{\prime}=H_{i}+\Delta H_{i}(\Delta x_{i}^{D})\;\Longrightarrow\;
\mathrm{Col}(H_{i})\neq\mathrm{Col}(H_{i}^{\prime})
\end{equation}
so any column-space attack vector
$c_{i}^{t}=H_{i}\,a_{S\text{-}FDI,i}^{t}$ built for the original
$H_{i}$ no longer lies in $\mathrm{Col}(H_{i}^{\prime})$, and the
residual grows above the chi-squared threshold with probability
increasing in the smallest principal angle between the two column
spaces~\cite{ref13}. The block $p_{BES,i}^{t}$ alters the apparent
power-injection signature the attacker may have profiled, while
$p_{tie,i}^{t}$ acts as a third MTD surface, since a sudden reversal
or magnitude change of $P_{ij}^{exch,t}$ disrupts any attacker model
assuming a quasi-static exchange schedule. The net export is realized
physically as a controllable load at a designated boundary bus, so
the tie-line action propagates through the AC-PF result.

\subsection{Multi-Agent PPO with Shared Encoder \& Private Heads}

Each ACA agent is trained with PPO in its multi-agent form
(MAPPO)~\cite{ref22}, preferred over off-policy alternatives for its
stability under the noisy rewards arising from FDI events and AC-PF
fallbacks, its compatibility with continuous bounded MTD actions, and
its clipped objective that prevents catastrophic policy jumps. Let
$\theta_{i}=(\Phi,\psi_{i})$ collect the shared encoder weights
$\Phi$ and the private actor/critic-head weights $\psi_{i}$. The
clipped surrogate objective is:
\begin{equation}\label{eq:clip}
L^{CLIP}(\theta_{i})=\E_{t}\!\Bigl[\min\bigl(\rho_{t}\hat{A}_{i,t},
\mathrm{clip}(\rho_{t},1{-}\epsilon_{c},1{+}\epsilon_{c})
\hat{A}_{i,t}\bigr)\Bigr]
\end{equation}
with importance ratio
$\rho_{t}=\pi_{\theta_{i}}(a_{i}^{t}|s_{i}^{t})/
\pi_{\theta_{i}^{\text{old}}}(a_{i}^{t}|s_{i}^{t})$. Advantages
$\hat{A}_{i,t}$ are obtained from GAE, and the total agent loss
combines the clipped policy term with value and entropy terms,
$\mathcal{L}^{\theta_{i}}=-L^{CLIP}+c_{v}L^{V}-c_{e}L^{H}$, with
coefficients $c_{v}$ and $c_{e}$.

\subsection{Federated Learning \& Heterogeneous Policy Design}

FL approach binds the $M$ MAPPO agents into a single
federation without exchanging raw observations, rewards, or
trajectories. At round $r$, the server broadcasts the global encoder
$\Phi^{r}$ and agent $i$ runs the local PPO update
$\Phi_{i}^{r+1}=\PPOLocalOp(\Phi^{r},\mathcal{T}_{i}^{r},E,\eta_{l}^{r})$
on its private rollouts and uploads only the encoder update
$\Delta\Phi_{i}^{r}=\Phi_{i}^{r+1}-\Phi^{r}$, the windowed mean
reward $R_{i}^{r}$, and the detection-quality score
$\eta_{q,i}^{r}$. The private actor and critic heads never leave
their host MG. Classical FedAvg would set
$\Phi^{r+1}=\Phi^{r}+\sum_{i}(n_{i}/n)\Delta\Phi_{i}^{r}$, but FedAvg
is provably unsafe under even a single adversarial
client~\cite{ref34}, so the server replaces it with the
Byzantine-resilient rule below.

Because MGs differ in their numbers of D-FACTS lines, BES units, and
tie-line neighbours, their action spaces have different
cardinalities. Forcing a common layout would either leak topology by
zero-padding to the worst case or restrict the federation to a
homogeneous subset. Instead, all MGs share an encoder $E_{\Phi}$
producing an embedding $h_{i}^{t}=E_{\Phi}(s_{i}^{t})$, while each MG
keeps a private actor head $\pi_{\psi_{i}}$ with
$a_{i}^{t}\sim\mathcal{N}(\mu_{i}^{t},\mathrm{diag}(\sigma_{i}^{2}))$
and a private critic head. Only $\Phi$ is federated; the
dimensionality and content of $\psi_{i}$ never appear in any uploaded
message, providing structural privacy on top of the
differential-privacy mechanism.

\subsection{Byzantine-Resilient Federated Aggregation}

An adversarial MG can attempt to derail the controller by submitting
an arbitrary update $\Delta\Phi_{i}^{r}=v_{i}$. BR-FedMAPPO uses a
two-stage robust rule. First, the server applies coordinate-wise
$\beta$-trimmed-mean filtering \cite{ref35}:
\begin{equation}\label{eq:tm}
\mathrm{TM}_{\beta}\bigl(\Delta\Phi^{r}\bigr)_{k}
=\frac{1}{(1-2\beta)M}\sum_{i\in\mathcal{Y}_{k}^{\beta}}\!\Delta\Phi_{i,k}^{r}
\end{equation}
For comparison, the Krum rule~\cite{ref36}
$K(\Delta\Phi^{r})=\arg\min_{i}\sum_{j\in\mathcal{N}_{i}^{k}}
\|\Delta\Phi_{i}^{r}-\Delta\Phi_{j}^{r}\|_{2}^{2}$ with $k=M-f-2$ is
also implemented. Second, we introduce the reward-weighted
aggregation coefficient that couples FL aggregation to closed-loop
grid behavior:
\begin{equation}\label{eq:rwcoef}
w_{i}^{r}=\frac{R_{i}^{r}\cdot\eta_{q,i}^{r}}
{\sum_{j\in\mathcal{S}_{bng}^{r}}R_{j}^{r}\cdot\eta_{q,j}^{r}}
\end{equation}
where $\eta_{q,i}^{r}=F_{1,i}^{r}\cdot(1-\mathrm{FPR}_{i}^{r})\in[0,1]$
is the detection-quality score, which collapses to near zero whenever
a client raises false alarms on most steps. The combined update is:
\begin{equation}\label{eq:rwupdate}
\Phi^{r+1}=\Phi^{r}+\sum_{i\in\mathcal{S}^{r}}w_{i}^{r}\,
\mathrm{TM}_{\beta}\bigl(\Delta\Phi_{i}^{r,DP}\bigr)
\end{equation}
Convergence under $f<M/4$ Byzantine clients follows
from~\cite{ref32}. The key advantage over coordinate-wise median is
that a chronically over-alarming MG is automatically de-weighted
through $\eta_{q,i}^{r}$, whereas median has no behavioural feedback.

\subsection{Differential Privacy Mechanism}

To complement the structural privacy of the heterogeneous
architecture, BR-FedMAPPO adds a Gaussian-mechanism
$(\epsilon,\delta)$-differential-privacy layer. Before transmission,
every client clips its encoder update to an $\ell_{2}$ norm $C$ and
adds isotropic Gaussian noise:
\begin{equation}\label{eq:dp_clip}
\Delta\Phi_{i}^{r,DP}=\Delta\Phi_{i}^{r}\cdot
\min\!\left(1,\frac{C}{\|\Delta\Phi_{i}^{r}\|_{2}}\right)+
\mathcal{N}\!\bigl(0,\sigma_{DP}^{2}\,\mathbf{I}\bigr)
\end{equation}
with $\sigma_{DP}=C\sqrt{2\ln(1.25/\delta)}/\epsilon$. The noise
interacts smoothly with the trimmed-mean filter: coordinate-wise
outliers are absorbed by the trim, while the unbiased Gaussian
preserves the expected aggregation value.

\subsection{Adaptive Islanding Mechanism}

The binary islanding indicator follows the hysteresis rule:
\begin{equation}\label{eq:island}
I_{i}^{iso,t}=\begin{cases}
1,& I_{i}^{iso,t-1}=0\ \wedge\ \alpha_{iso,i}^{t}>\tau_{iso},\\[2pt]
1,& I_{i}^{iso,t-1}=1\ \wedge\ (\tau_{i}^{hold,t}>0
        \ \vee\ \alpha_{iso,i}^{t}>\tau_{iso}),\\[2pt]
0,& \text{otherwise}
\end{cases}
\end{equation}
To prevent rapid switching oscillations, the hold counter $\tau_{i}^{hold,t}$ enforces a minimum isolation duration of $T_{iso,\min}$. During this period, the external breaker opens, a local slack generator is designated, tie-line exchanges cease, and proportional load-shedding is applied as $\Ld_{i}^{t}\!\leftarrow\!\Ld_{i}^{t}\cdot\min(1,S_{i}^{t}/D_{i}^{t})$, where $S_{i}^{t}$ and $D_{i}^{t}$ denote local supply and demand. Furthermore, an islanding event tracker systematically logs critical pre, during, and post-isolation snapshots, including power balance, voltage profiles, and cost shifts, for both the engaged MG and its neighbours. These comprehensive records, alongside the event duration $\Delta t_{iso,i}$, fundamentally support the propagation analysis detailed in Section~\ref{sec:results}.

\subsection{Overall Training Procedure}

The complete procedure is summarized in
Algorithm~\ref{alg:brfedmappo} and alternates between local PPO
updates and Byzantine-resilient server aggregation.

\begin{algorithm}[!t]
\caption{BR-FedMAPPO Training Procedure}
\label{alg:brfedmappo}
\scriptsize
\begin{algorithmic}[1]
\Require $M$ MGs; $R{=}100$ federation rounds; $E_{ep}$ episodes/round;
         $K{=}48$ control steps/episode; batch $B{=}64$; PPO epochs $=10$;
         attack probability $p_{atk}{=}0.45$; DP std
         $\sigma_{DP}{=}0.01$; trim ratio $\beta{=}0.25$.
\Ensure Global encoder $\Phi^{R}$; local heads $\{\psi_{i}^{R}\}$;
        deployed MTD schedules; detection diagnostics.
\State Initialize $\Phi^{0},\{\psi_{i}^{0},V_{\psi_{i}}^{0}\}_{i=1}^{M}$;
       reset $E_{i,k}\!\leftarrow\!0.5\,E_{\max}$.
\State \textbf{Warm-up:} run 200 steps with random uniform (75\%) +
       zero (25\%) actions; train AE; freeze chi-squared,
       cumulative-sum, magnitude and AE thresholds at the empirical
       99th percentile.
\For{$r=0,1,\ldots,R-1$} \Comment{federation round}
   \State Server broadcasts $\Phi^{r}$ to all MGs.
   \For{each MG $i\in\mathcal{F}$ \textbf{in parallel}}
      \State Load $\Phi^{r}$ into local encoder $E_{\Phi}$;
             snapshot $\{TP,FP,FN,TN\}$.
      \For{$\text{episode}=1,\ldots,E_{ep}$}
         \State Reset MG~$i$ state; restore baseline loads.
         \For{$k=0,\ldots,K-1$}
            \State Sample
                  $a_{i}^{t}\sim\pi_{\theta_{i}}(\cdot|s_{i}^{t})$ as
                  in \eqref{eq:action}.
            \State Apply D-FACTS:
                  $x_{l}\!\leftarrow\!x_{l}^{base}+\Delta x_{i,l}^{D}$.
            \State Apply BES; update $E_{i,k}^{t+1}$ via \eqref{eq:bes}.
            \State Apply islanding with hysteresis~\eqref{eq:island}.
            \State Settle tie-line:
                  $P_{ij}^{exch}\!\leftarrow\!\tfrac{1}{2}(p_{ij}-p_{ji})$;
                  realize as load at boundary bus.
            \State Decide FDI: if burst active, continue; else
                  $\sim\!\mathrm{Bern}(p_{atk})$.
            \State Run AC-PF; collect $z_{i}^{t}$.
            \State Inject FDI: $\tilde{z}_{i}^{t}=z_{i}^{t}+c_{i}^{t}$
                  per~\eqref{eq:fdi}.
            \State Run 4-statistic detector;
                  $d_{i}^{t}\!=\!\indic[\sum\text{flags}\ge\tau_{vote}]$.
            \State Compute $r_{i}^{t}$ per~\eqref{eq:reward}; push
                  transition to rollout buffer.
         \EndFor
      \EndFor
      \State Compute GAE; PPO update for 10 epochs, minibatch $B{=}64$.
      \State Compute $R_{i}^{r}$ and
             $\eta_{q,i}^{r}=F_{1,i}^{r}(1-\mathrm{FPR}_{i}^{r})$.
      \State Encoder update
             $\Delta\Phi_{i}^{r}$; clip to $\|\cdot\|_{2}\le C$; add
             DP noise~\eqref{eq:dp_clip}.
      \State Send $(\Delta\Phi_{i}^{r,DP},R_{i}^{r},\eta_{q,i}^{r})$ to
             server.
   \EndFor
   \State \textbf{Server:} apply trimmed mean~\eqref{eq:tm}; compute
          $w_{i}^{r}$~\eqref{eq:rwcoef}; update
          $\Phi^{r+1}$~\eqref{eq:rwupdate}; update DP accountant.
\EndFor
\State \Return $\Phi^{R},\{\psi_{i}^{R}\}$; detection summary,
       training metrics, confusion-matrix evolution, and aggregation
       diagnostics.
\end{algorithmic}
\end{algorithm}

\section{Experimental Setup}\label{sec:setup}

The cluster contains four interconnected MGs on both test
systems, $M=4$. On the small-scale test system each MG is an IEEE
30-bus network with 41 lines, equipped with three BES units placed
at buses 10, 20, and 25 with rated energies of 10, 12, and 8~MWh
respectively, and a per-unit power limit of 3~MW. The number of
D-FACTS-equipped lines is heterogeneous across MGs, namely 5, 6, 4,
and 5 for MG1 through MG4 respectively, for a cluster total of
twenty D-FACTS devices. The boundary bus through which the net
inter-MG export is imposed as a controllable load is bus~22. On the
large-scale test bed each MG is an IEEE 118-bus network with 173
lines and 53 generators, equipped with six BES units per MG and
10--12 D-FACTS-equipped lines per MG; the boundary bus is bus~44.
The federated topology connects MG1 to MG2 and MG3, MG2 to MG1,
MG3, and MG4, MG3 to MG1, MG2, and MG4, and MG4 to MG2 and MG3.
The voltage band is $V\in[0.94,1.06]$~p.u.\ on both test systems. The PPO learning rates are $3\times10^{-4}$ for the actor and
$10^{-3}$ for the critic. The discount and GAE factors are
$\gamma=0.99$ and $\lambda=0.95$, the clipping ratio is
$\epsilon_{clip}=0.20$, the entropy coefficient is $c_{e}=0.01$,
the value coefficient is $c_{v}=0.50$, the gradient norm is clipped
at $0.50$, the PPO epoch count is $K_{ppo}=10$, and the mini-batch
size is $64$. The federation runs $R=100$ rounds with the warm-up phase of $W=200$ steps
on the 30-bus test system and $W=280$ steps on the 118-bus test system.

\section{Simulation Results and Discussion}\label{sec:results}

This section presents a comprehensive evaluation of the proposed framework across both test beds. We initially establish the operational advantages of the federated architecture by comparing it with a fully decentralized baseline. Subsequently, we validate the robustness of various aggregation rules in hostile environments and examine the adaptive islanding mechanism for effective attack containment. The analysis concludes by assessing the overall defense performance against cyber-physical intrusions, detailing a sensitivity study of critical system parameters, and evaluating the balance among privacy, utility, and computational overhead.

\subsection{A Comparative Evaluation of Federated \& Decentralized Approaches}
This subsection evaluates the operational advantages of the federated architecture over fully decentralized training. The defensible gain appears in detection sensitivity and transfer.
Federation lifts mean recall from $54.56\%$ to $55.28\%$ on the
118-bus cluster, $+1.33\%$ relative, and from $54.20\%$ to $54.81\%$
on the 30-bus system cluster, while the cross-MG generalization study shows
an FL policy evaluated on a MG it never trained on losing only
$1.61$ recall points for 118-bus, and $0.89$ points for 30-bus, relative to
the same-MG case, at zero false positives on every transfer
pair. A decentralized policy has no comparable transfer object because its
action head and encoder are entangled and cannot be moved. The shared
encoder is therefore not a marginal accuracy tweak but the only
component in the design that survives a topology change, which is
exactly the property an operator needs when a new MG joins the
cluster. This sensitivity gain is not free, and the trade-off is itself
informative. On 118-bus the federated detector raises recall but also
raises the mean false-positive rate, from $0.088$ to $0.142$, and
widens the spread of recall across the four MGs rather than
narrowing it. 
\begin{table}[!t]
\caption{FL \& Non-FL results in terms of Recall and transfer improvement.}
\label{tab:flnofl_real}
\centering
\scriptsize
\renewcommand{\arraystretch}{0.95}
\setlength{\tabcolsep}{3.0pt}
\begin{tabular}{@{}lcccc@{}}
\toprule
\textbf{Metric} & \multicolumn{2}{c}{\textbf{IEEE 30-bus}}
                & \multicolumn{2}{c}{\textbf{IEEE 118-bus}} \\
\cmidrule(lr){2-3}\cmidrule(lr){4-5}
        & Non-FL & FL & Non-FL & FL \\
\midrule
Recall (\%)             & 54.20 & 54.81 & 54.56 & 55.28 \\
Detection quality       & 0.696 & 0.698 & 0.646 & 0.621 \\
Mean FPR                & 0.013 & 0.014 & 0.088 & 0.142 \\
Cross-MG recall drop (pp) & 1.17 & 0.89 & 5.36 & 1.61 \\ 
\bottomrule
\end{tabular}
\end{table}

\begin{figure}[!t]
\centering
\includegraphics[width=\columnwidth]{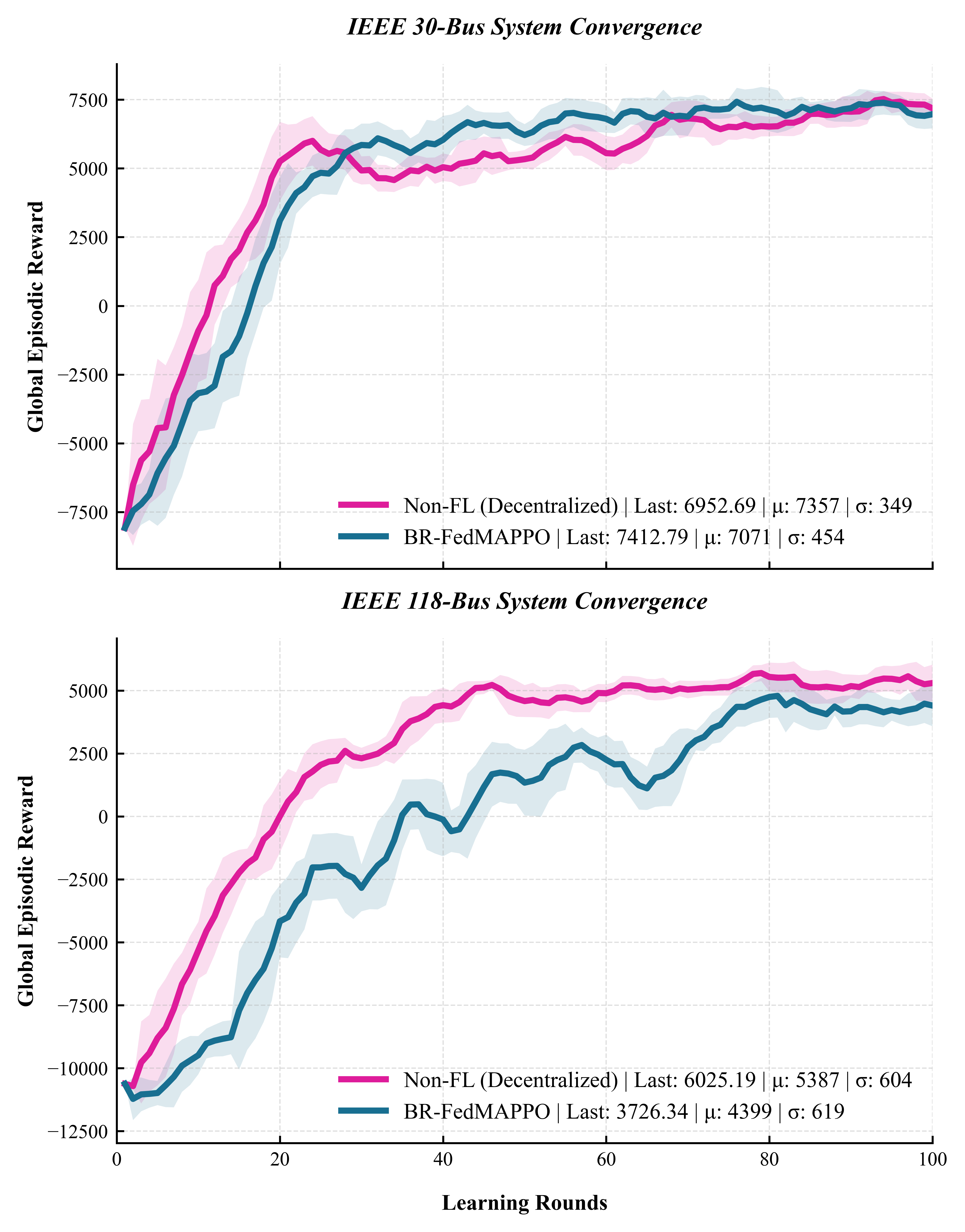}
\caption{Episodic average reward of four MGs over a hundred rounds for the proposed BR-FedMAPPO and baseline approaches for both test systems.}
\label{fig:detrate_quality}
\end{figure}

\subsection{Aggregation under Byzantine Poisoning}

\begin{figure}[!t]
\centering
\includegraphics[width=\columnwidth]{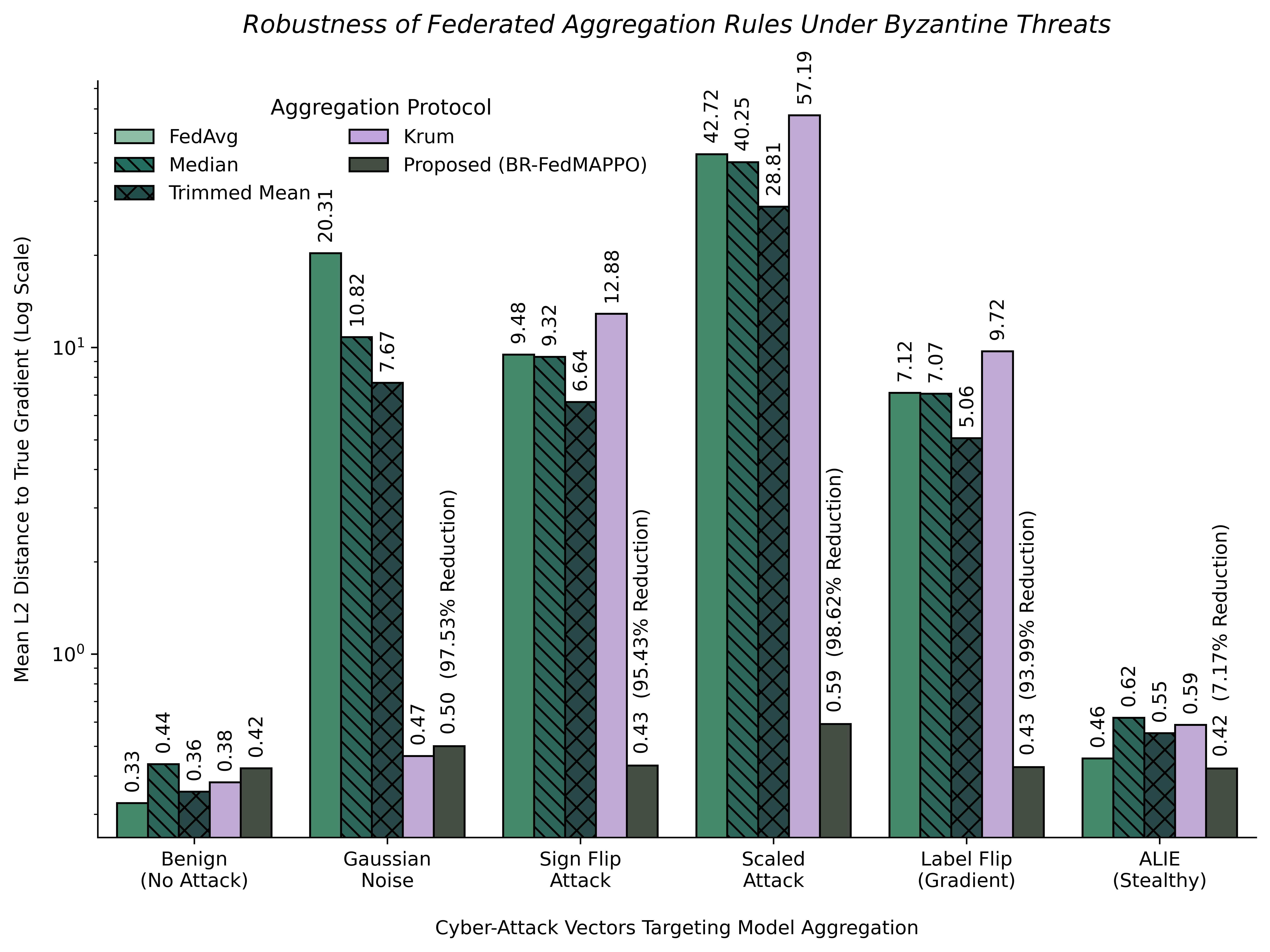}
\caption{Robustness of the federated aggregation rules under
Byzantine conditions.}
\label{fig:bzt}
\end{figure}

The second study benchmarks the proposed reward-weighted rule against established aggregation mechanisms including federated averaging, median, trimmed mean, and Krum. Table \ref{tab:agg} and Fig. \ref{fig:bzt} detail these comparative outcomes. The vulnerability of standard aggregation methods in hostile environments stems from their indiscriminate incorporation of poisoned updates. While conventional robust filters like Krum or trimmed mean successfully reject extreme outliers, they fundamentally lack closed-loop behavioral awareness. The proposed reward-weighted aggregation comprehensively overcomes this limitation by dynamically evaluating the operational efficacy of each participant. By integrating the detection quality score directly into the aggregation logic, the server actively penalizes any MG that exhibits compromised behavior or chronically generates false alarms. This structural robustness is explicitly validated through a direct Byzantine poisoning experiment. 

\begin{table}[!t]
\caption{Aggregation-Method Benchmark on the IEEE 30-bus and IEEE 118-bus Test Systems.}
\label{tab:agg}
\centering
\renewcommand{\arraystretch}{1.15}
\setlength{\tabcolsep}{3pt}
\footnotesize
\begin{tabular}{|l|c|c|c|c|c|c|}
\hline
\multirow{2}{*}{\textbf{Method}} &
\multicolumn{2}{c|}{\textbf{Reward}} &
\multicolumn{2}{c|}{\textbf{Recall (\%)}} &
\multicolumn{2}{c|}{\textbf{Detection Quality}}\\
\cline{2-7}
 & 118-bus & 30-bus & 118-bus & 30-bus & 118-bus & 30-bus \\
\hline
FedAvg          & 1{,}083.2 & 1{,}797.3 & 56.5 & 55.3 & 0.638 & 0.701 \\
Median          & 929.2     & 1{,}783.3 & 59.0 & 54.9 & 0.610 & 0.697 \\
Trimmed mean    & 1{,}099.7 & 1{,}767.7 & 57.1 & 54.7 & 0.623 & 0.698 \\
Krum            & 1{,}114.3 & 1{,}818.1 & 57.1 & 55.0 & 0.635 & 0.698 \\
Reward-weighted & 979.5     & 1{,}807.3 & 58.4 & 54.9 & 0.637 & 0.698 \\
\hline
\end{tabular}
\end{table}

As illustrated in Fig. \ref{fig:bzt}, explicit malicious injections generate enormous $\ell_2$ deviations from the actual reference gradient. Specifically, scaled and Gaussian poisoning vectors sit roughly 130.79 and 62.18 times above the benign update norm, while sign-flip and gradient-label-flip attacks exhibit deviations of 29.03 and 21.79 times the baseline. Under these severe conditions, baseline approaches like federated averaging experience catastrophic failure and yield massive aggregation errors. In stark contrast, the proposed reward-centric algorithm precisely identifies and isolates malicious clients, achieving a significant 98.62\% reduction in aggregation error. 
\begin{figure*}[!t]
\centering
\includegraphics[width=0.95\textwidth]{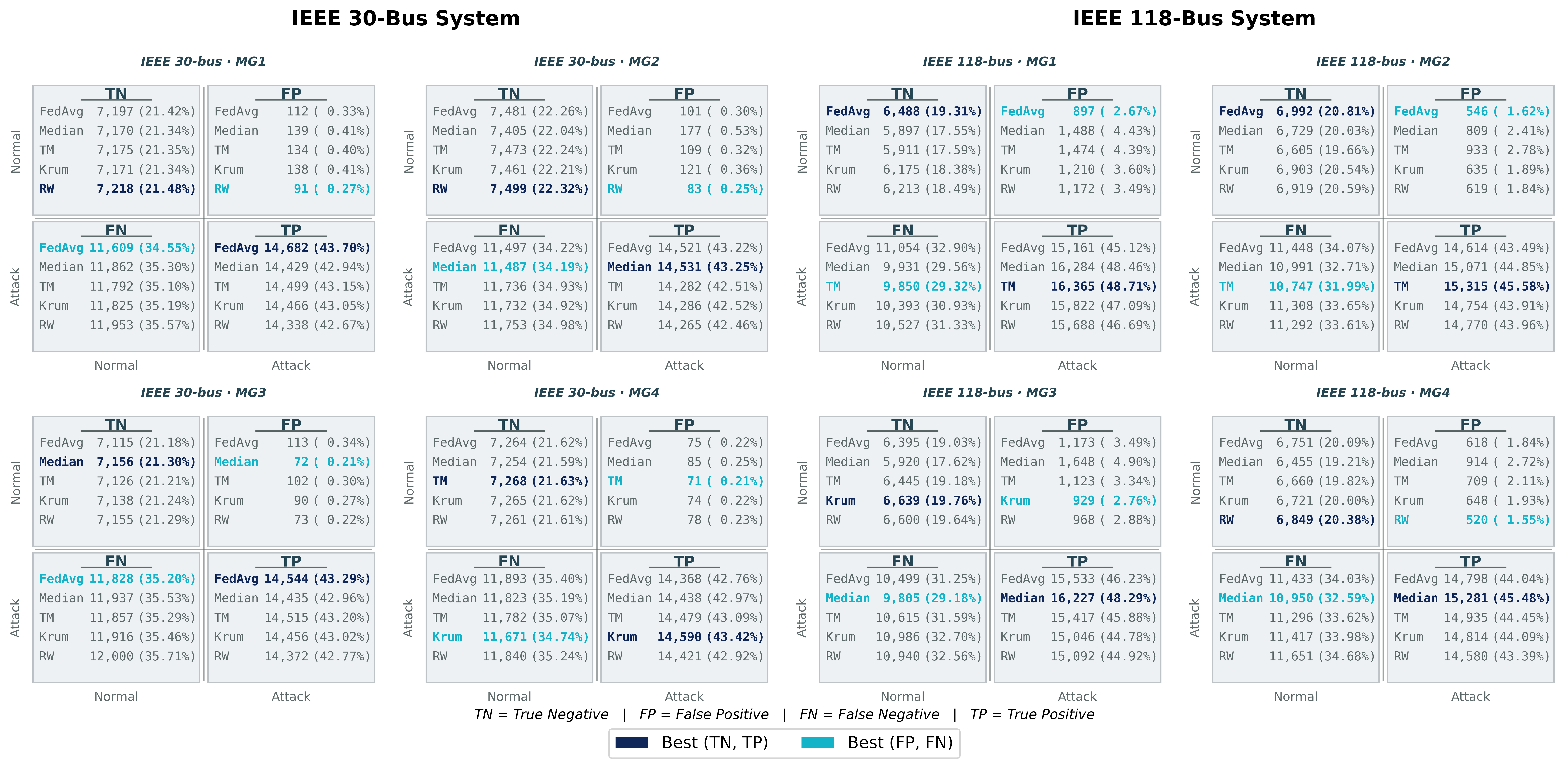}
\caption{Detailed per-MG confusion matrix decomposition comparing the five federated aggregation rules across the IEEE 30-bus and 118-bus systems.}
\label{fig:confusion}
\end{figure*}


To reveal the underlying behavior of the aggregate metrics, Fig. \ref{fig:confusion} presents a detailed per-MG confusion matrix decomposition. The results highlight a critical operational trade-off carried out by the proposed BR-FedMAPPO model. While standard robust filters like Trimmed Mean and Median aggressively maximize True Positives on the complex 118-bus system, they inadvertently inflate the False Positive rates. In contrast, the proposed Reward-Weighted rule consistently excels at minimizing false alarms. For instance, RW achieves the absolute lowest FP counts in MG1 and MG2 of the 30-bus system, 91 and 83 false alarms, respectively. More importantly, on the highly non-linear 118-bus system, RW significantly suppresses false alarms in critical areas such as MG4, recording only 520 FPs compared to 914 for Median. This explicitly validates our core design claim: by integrating the detection-quality score into the aggregation weights, the server successfully penalizes chronically over-alarming agents, thereby prioritizing system stability and minimizing costly precautionary islanding events without severely degrading attack recall.

\subsection{Adaptive-Islanding Behaviour}

The third study demonstrates that adaptive islanding operates as an intelligent and attack-correlated control action rather than a simple reactive reflex. In the 118-bus system, 77.8\% of the islanding events coincide with genuinely active attacks as detailed in Table \ref{tab:island_real}. The remaining instances are precautionary disconnections driven by elevated anomaly scores near the detection threshold. Furthermore, the duration of these events is heavily concentrated around the four-step minimum hold limit with a median of 4 and a mean of approximately 4.08. This explicitly confirms that the hysteresis mechanism defined in \eqref{eq:island} successfully mitigates the rapid switching oscillations that typically plague conventional threshold-based schemes. The containment claim is supported by the neighbour-response tracker.
When an MG engages islanding, a physical neighbour is co-islanded only
$20.3\%$ of the time, and neighbour minimum voltage holds at $0.943$~ per unit, essentially at the lower band edge but never collapsing, while the
neighbour absorbs only a sub-MW tie-line redirection. In other words,
opening one breaker does not cascade: the corrupted-measurement path is
cut at the tie-line, the disturbance is converted into a small,
bounded voltage and power-flow adjustment next door, and the cluster
does not unravel. 

Federation increases islanding activity during genuine attack occurrences. The FL controller closes
$65\%$ more events than its non-federated counterpart on the 118-bus system, $4{,}649$ compared to \ $2{,}811$, with a corresponding $66.15\%$ rise in load-shedding
energy. The FL policy detects more attacks, higher recall, and therefore acts more often; the attack-triggered
fraction is in fact slightly higher under FL, $77.8\%$ against\
$76.6\%$, so the additional engagements are predominantly justified
rather than spurious. The correct framing is that load-shedding here is a controlled defensive expenditure that scales with detection
sensitivity, not an efficiency loss, the per-event duration is
unchanged, so the system is shedding for the same short, bounded
intervals, just on more confirmed threats. Additionally, the baseline operational cost per MG is approximately \$64266.94, and the activation of the defensive isolation mechanism incurs a fundamentally negligible 1.66\% increase in the aggregate operational costs of the neighboring networks.

\begin{table}[!t]
\caption{Adaptive-islanding lifecycle on the IEEE 118-bus cluster per-run aggregate over four MGs.}
\label{tab:island_real}
\centering
\scriptsize
\renewcommand{\arraystretch}{1.1}
\setlength{\tabcolsep}{4.0pt}
\begin{tabular}{@{}lccc@{}}
\toprule
\textbf{Statistic} & \textbf{Non-FL} & \textbf{FL} & \textbf{$\Delta$} \\
\midrule
Total closed events              & 2{,}811 & 4{,}649 & $+65.4\%$ \\
Attack-triggered fraction (\%)   & 76.6    & 77.8    & $+1.2$ pp \\
Mean duration (steps)            & 4.06    & 4.08    & $+0.02$ \\
Median duration (steps)          & 4       & 4       & $0$ \\
Total shed energy (MWh)          & 1.79    & 2.97    & $+66.15\%$ 
\\
Neighbour co-island rate (\%)    & ---     & 20.3    & --- \\
Neighbour $V_{\min}$ (p.u.)      & ---     & 0.943   & --- \\
Mean neighbour cost shift (\$/h) & ---     & 3199.66    & --- \\
\bottomrule
\end{tabular}
\end{table}

\begin{table}[!t]
\caption{Relative performance variation in percent of the FL framework compared to the Non-FL baseline.}
\label{tab:fl_improvements}
\centering
\scriptsize
\renewcommand{\arraystretch}{1.15}
\setlength{\tabcolsep}{3.5pt}
\resizebox{\columnwidth}{!}{%
\begin{tabular}{@{}lrrrrrrrr@{}}
\toprule
\multirow{2}{*}{\textbf{Metric}} & \multicolumn{4}{c}{\textbf{IEEE 30-bus Test System}} & \multicolumn{4}{c}{\textbf{IEEE 118-bus Test System}} \\
\cmidrule(lr){2-5} \cmidrule(l){6-9}
 & \textbf{MG1} & \textbf{MG2} & \textbf{MG3} & \textbf{MG4} & \textbf{MG1} & \textbf{MG2} & \textbf{MG3} & \textbf{MG4} \\
\midrule
$\Delta C_i$ (\%) & -39.33 & 9.31  & -5.71 & -9.28  & -10.60 & -12.34 & -8.59 & -1.88 \\
$\Delta P_{shed, i}$ (MW) & 0.11  & -0.03 & 0.02  & -0.09 & 0.59 & 0.25  & 0.16 & 0.16 \\
$\Delta \alpha_{iso, i}$ (\%)      & 32.51  & -8.76 & 6.90  & -24.90 & 129.10 & 63.51  & 27.48 & 50.20 \\
Recall Improvement (\%)          & 1.47   & 0.43  & 0.86  & 1.74   & 3.89   & 3.81   & 4.60  & 0.70 \\
\bottomrule
\end{tabular}%
}
\vspace{-2mm}
\end{table}

\subsection{FDI-Detection Performance}

The FL framework induces significant operational shifts across the distributed network as detailed in Table \ref{tab:fl_improvements}. Beyond these relative variations, the absolute detection performance reflects a highly asymmetric profile entirely driven by attack class. Specifically, the framework achieves a mean precision of 0.993 and 0.937 for the 30-bus and 118-bus test systems respectively, while strictly maintaining a mean false-positive rate at 0.0135 for the 30-bus configuration and 0.142 for the 118-bus system. While conventional static intrusion detection systems inherently struggle with sophisticated manipulations, the proposed architecture decisively neutralizes them through its MTD mechanism. By dynamically altering line reactances, the framework continuously shifts the active Jacobian matrix. Consequently, stealthy attack vectors synthesized against an outdated static topology fundamentally mismatch the current physical state, generating large residual errors and enabling high stealthy-attack recall across all MGs in both test systems.

Conversely, the reduced step-wise sensitivity to sparse attacks represents a deliberate optimization outcome rather than a structural deficiency. Sparse injections operate independently of physical network equations, perturbing a limited number of channels with magnitudes closely resembling ambient measurement noise. Because the DRL formulation imposes stringent penalties for false alarms, the ACA agent rationally learns to tolerate these low-impact anomalies. This behavioral adaptation results in sparse step recalls of [0.101, 0.101, 0.088, 0.081] for the 30-bus test bed and [0.108, 0.036, 0.082, 0.063] for the 118-bus test systems. Elevating the detection sensitivity to capture every minor sparse injection would inevitably misclassify natural operational noise and severely undermine system stability. Instead, the framework relies on the cumulative-sum statistic to successfully identify sustained events, achieving robust sparse burst recalls of [0.962, 0.965, 0.958, 0.961] and [0.968, 0.952, 0.960, 0.955] for the 30-bus and 118-bus configurations respectively. Ultimately, this inverted detection profile mathematically proves that the physical defense layer effectively eliminates stealthy vulnerabilities, while the intelligent learning policy establishes an optimal equilibrium between strict detection precision and continuous operational stability.


\subsection{Sensitivity Analysis}

\begin{figure}[!t]
\centering
\includegraphics[width=\columnwidth]{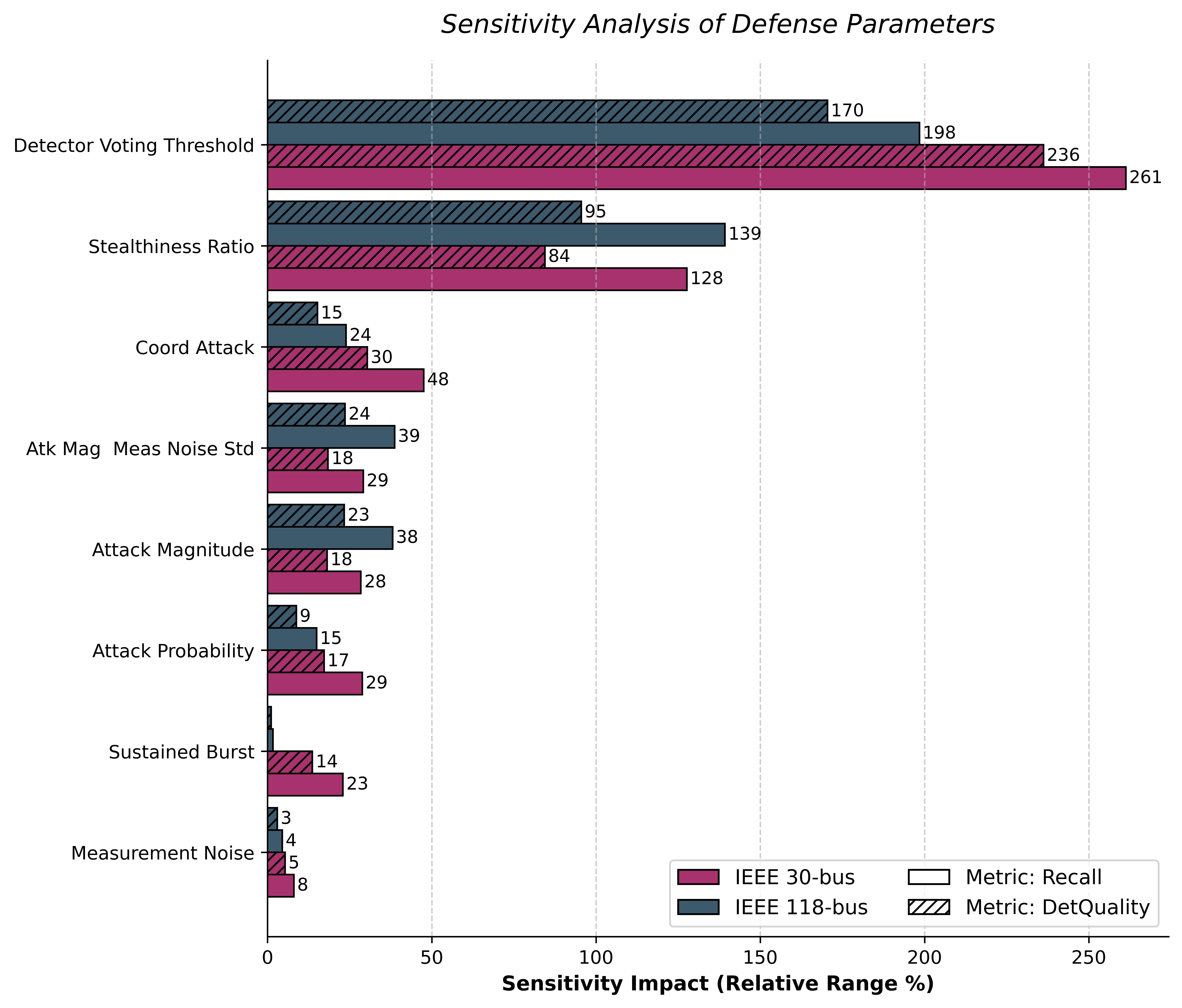}
\caption{Sensitivity of detection quality and recall to eight
defense parameters on the IEEE 30-bus and IEEE 118-bus systems,
ranked by relative range.}
\label{fig:tornado}
\end{figure}

The sensitivity analysis evaluates the operational robustness of the trained policy across varying deployment parameters. As illustrated in Fig. \ref{fig:tornado}, the detector voting threshold and attack stealthiness ratio are the primary factors influencing detection quality. The voting threshold serves as a critical balance: relaxing it inflates false alarms, whereas overly strict criteria blind the system to sophisticated intrusions. Similarly, attack stealthiness dictates how closely malicious injections mimic valid physical states, directly challenging the anomaly filters. In contrast, the framework exhibits remarkable immunity to ambient measurement noise. This resilience stems from the data-driven calibration phase, which uses adaptive historical percentiles to absorb baseline environmental variance seamlessly without inflating the false positive rate. 
\section{Conclusion}\label{sec:conclusion}

This paper presented BR-FedMAPPO, a Byzantine-resilient federated multi-agent reinforcement learning framework designed to secure interconnected microgrids against sophisticated cyber-physical manipulations. By integrating a triple-surface Moving Target Defense (MTD) with a learned adaptive islanding mechanism, the proposed architecture offers a robust, privacy-preserving defense strategy. The quantitative evaluation across the IEEE 30-bus and 118-bus test beds yields four primary conclusions. First, the federated architecture improves structural privacy and cross-MG transferability while maintaining competitive operational performance. The FL framework inherently shields local defense topologies and reduces the average operational cost increment by up to $39.33\%$ in the 30-bus network and $12.34\%$ in the 118-bus network, proving that collaborative learning enhances cost-aware dispatch under threat conditions. Second, the shared-encoder design ensures exceptional cross-network generalization during adaptive isolation or topology changes. When subjected to cross-MG transfer evaluation, the FL policy experiences a marginal recall drop of only $0.89$ percentage points on the 30-bus system and $1.61$ percentage points on the 118-bus system. In contrast, the non-federated approach suffers significantly higher degradation, drops of $1.17$ and $5.36$ percentage points, respectively, highlighting the necessity of representation sharing for resilient autonomous isolation. 
Third, the dynamic triple-surface MTD mechanism effectively neutralizes stealthy intrusions. By continuously shifting line reactances, BES dispatch, and tie-line exchanges, the defense framework forces pre-calculated attack vectors out of the valid operational space, achieving at least 95\% burst recall against stealthy FDI attacks across all microgrids while maintaining a false positive rate below $0.142$. Finally, the proposed reward-weighted aggregation rule exhibits firm stability in hostile federated environments. By dynamically scaling participant influence based on real-time detection quality, the algorithm achieves a $98.62\%$ reduction in aggregation error under severe Byzantine poisoning compared to standard federated averaging. 

\bibliographystyle{IEEEtran}
\bibliography{references}

\end{document}